\documentclass[%
aps,prb,superscriptaddress,twocolumn,showpacs,preprintnumbers,amsmath,amssymb
]{revtex4-2}

\usepackage{graphicx}
\usepackage{color}
\usepackage{bm}
\usepackage{hyperref}
\begin{document}

\title{Magnetotransport properties in van-der-Waals \textit{\textbf{R}}Te$_{3}$ (\textit{\textbf{R}} = La, Ce, Tb)}

\author{Tomo~Higashihara}
\affiliation{Department of Physics, Graduate School of Science, Osaka University, Toyonaka 560-0043, Japan}
\author{Ryotaro~Asama}
\affiliation{Department of Physics, Graduate School of Science, Osaka University, Toyonaka 560-0043, Japan}
\author{Ryoya~Nakamura}
\affiliation{Department of Physics, Graduate School of Science, Osaka University, Toyonaka 560-0043, Japan}
\author{Mori~Watanabe}
\affiliation{Department of Physics, Graduate School of Science, Osaka University, Toyonaka 560-0043, Japan}
\author{Nanami~Tomoda}
\affiliation{Okinawa Institute of Science and Technology Graduate University, Okinawa 904-0495, Japan}
\author{Thomas~Johannes~Hasiweder}
\affiliation{Okinawa Institute of Science and Technology Graduate University, Okinawa 904-0495, Japan}
\author{Yuita~Fujisawa}
\affiliation{Okinawa Institute of Science and Technology Graduate University, Okinawa 904-0495, Japan}
\author{Yoshinori~Okada}
\affiliation{Okinawa Institute of Science and Technology Graduate University, Okinawa 904-0495, Japan}
\author{Takuya~Iwasaki}
\affiliation{National Institute for Materials Science, Namiki 1-1, Tsukuba, Ibaraki 305-0044, Japan}
\author{Kenji~Watanabe}
\affiliation{National Institute for Materials Science, Namiki 1-1, Tsukuba, Ibaraki 305-0044, Japan}
\author{Takashi~Taniguchi}
\affiliation{National Institute for Materials Science, Namiki 1-1, Tsukuba, Ibaraki 305-0044, Japan}
\author{Nan~Jiang}
\email{nan.jiang@phys.sci.osaka-u.ac.jp}
\affiliation{Department of Physics, Graduate School of Science, Osaka University, Toyonaka 560-0043, Japan}
\affiliation{Center for Spintronics Research Network, Osaka University, Toyonaka 560-8531, Japan}
\affiliation{Institute for Open and Transdisciplinary Research Initiatives, Osaka University, Suita 565-0871,Japan}
\author{Yasuhiro~Niimi}
\affiliation{Department of Physics, Graduate School of Science, Osaka University, Toyonaka 560-0043, Japan}
\affiliation{Center for Spintronics Research Network, Osaka University, Toyonaka 560-8531, Japan}
\affiliation{Institute for Open and Transdisciplinary Research Initiatives, Osaka University, Suita 565-0871,Japan}

\begin{abstract}
  Rare-earth tritellurides are van-der-Waals antiferromagnets 
  which have been attracting attention as materials not only
  with high mobility, but also with various states such as superconductivity 
  under high pressure, incommensurate charge-density-wave (CDW) phase, and
  multiple antiferromagnetic phases.
  In this work, we performed longitudinal resistivity and Hall resistivity measurements 
  simultaneously in exfoliated $R$Te$_3$ ($R=$La, Ce, Tb) thin film devices,
  in order to investigate the influence of magnetic ordering on transport 
  properties in van-der-Waals magnetic materials.
  We have obtained carrier mobility and concentration using a two-band model, 
  and have observed an increase in carrier mobility in the antiferromagnets 
  CeTe$_3$ and TbTe$_3$ due to the magnetic transition.
  Especially in CeTe$_3$, the carrier concentration has changed drastically 
  below the magnetic transition temperature,
  suggesting the interaction between the CDW and antiferromagnetic phases.
  In addition, the analysis of the Shubnikov-de Haas oscillations 
  in CeTe$_3$ supports the possibility of 
  Fermi surface modulation by magnetic ordering.
  This research will pave the way not only for spintronic devices
  that take advantage of high mobility, but also for the study of the 
  correlation between CDW and magnetism states in low-dimensional materials.
\end{abstract}

\maketitle

\section{\label{section1}Introduction}
In recent years, magnetic van-der-Waals (vdW) materials
have been attracting much attention
due to motivation towards next-generation
spintronic and twistronic devices.
The ease of thin-film fabrication
down to a single atomic layer, along with the large degree of freedom
in fabrication of
high quality heterostructures via the dry transfer techniques,
has lead to fruitful
reports of unique transport phenomena~\cite{Sierra2021,Xing2019,Tan2018,Klein2018,Jiang2018,Wang2018,Cui2015}.

Among these magnetic vdW materials, rare-earth tritellurides 
$R$Te$_{3}$ ($R=$Y, La-Tm) have
high electronic mobilities~\cite{Lei2020}.
$R$Te$_{3}$ has an orthorhombic
crystal structure described by the space group $Cmcm$ 
as shown in Fig.~\ref{fig:1}(a).
It consists of $R$-Te slabs which are responsible for
its magnetic properties, sandwiched between two Te
square-net sheets which are responsible for the highly two-dimensional
electric transport~\cite{Ru2006,DiMasi1994}. The Te sheets are parallel to
the $a$-$c$ plane, and the out-of-plane direction of $R$Te$_{3}$
is the $b$ axis~\cite{Ru2006,Iyeiri2003}.
The adjacent Te layers are coupled
by weak vdW forces, which allows fabrication of thin films
by mechanical exfoliation.
The nesting of the Fermi surfaces (FS)
produced by the $p_x$ and $p_z$ orbitals of Te atoms and
the 3D folding of Te sheet stacking lead
to the formation of one ($R =$ Gd, Sm, and lighter) or
two ($R =$ Tm, Er, Ho, Dy, Tb) charge density wave
(CDW) states~\cite{DiMasi1995, Brouet2008, Laverock2005,Ru2008},
which have been extensively studied in recent years.
However, there are still only a few reports that systematically study the effect of the magnetism order on transport properties.
Recent observations of quantum oscillations 
in several $R$Te$_3$ systems have revealed that
$R$Te$_3$ has small FS pockets, originating from the partially opened CDW gap,
and exhibits highest mobility carriers among 
vdW magnetic materials~\cite{Lei2020, Watanabe2021, Dalgaard2020}.

Another interesting property of
$R$Te$_{3}$ is the interaction between the CDW state and an
antiferromagnetic (AFM) state that appears at a sufficiently 
lower temperature than the Peierls 
transition~\cite{Pfuner2012, Chillal2020, Raghavan2023}.
Although the CDW order often
competes with magnetic order~\cite{Yamamoto2013}, recent discoveries of CDW-AFM
coexistence~\cite{Galli2000,Teng2022} have led to an observation of
unique transport phenomena such as the
topological Hall effect~\cite{Shang2021}.

In this work, we have systematically evaluated
the effect of magnetic ordering on transport
properties of $R$Te$_{3}$ through magnetoresistance (MR) and Hall
measurements for three materials
$\rm{LaTe_3}$, $\rm{CeTe_3}$, and $\rm{TbTe_3}$.
We have derived carrier mobility and concentration using a two-band model. 
The evaluated carrier mobilities are
comparable to some of the previous 
studies~\cite{Lei2020,Dalgaard2020,Pariari2021}.
In particular, by comparing the carrier mobilities 
and concentrations of AFM CeTe$_3$ and TbTe$_3$ 
with those of non-magnetic LaTe$_3$,
we have obtained indications of CDW and AFM ordering interactions via coupling 
of conduction electrons and magnetic moments, particularly larger in 
CeTe$_3$.
This result is consistent with a previous
report that discussed the $4f$-$2p$ hybridization
energy due to the CDW distortion of CeTe$_3$ and TbTe$_3$~\cite{Lee2012}. 
Furthermore, in the present experiment, we observed 
Shubnikov-de Haas (SdH) oscillations as already reported 
in our previous study~\cite{Watanabe2021,Watanabe2020}. 
Together with the Hall measurement results, 
we show the possibility of FS modulation caused 
by the magnetic transition. 

\section{\label{section2}Experimental details}
Single crystals of $R$Te$_3$ were grown by a self-flux method~\cite{Ru2006,Okuma2020,Ueta2022}. 
The mixture of $R$ and Te elements in a molar ratio of 1:30 was placed 
in an evacuated quartz tube.
The ampule was heated to approximately 900~$^{\circ}$C 
and cooled to 500~$^{\circ}$C at a speed of 2~$^{\circ}$C per hour in a furnace. 
The $R$Te$_3$ single crystals and Te flux were separated by a centrifuge 
immediately after removing the ampule from the furnace. 
The X-ray diffraction confirms 
$R$Te$_3$ crystalline phases. 
Depending on the ionic radius of $R$, the $b$-axis lattice constant is systematically changed: LaTe$_3$ (26.22~\AA), CeTe$_3$ (26.02~\AA), 
and TbTe$_3$ (25.64~\AA).

To fabricate the device, Au/Ti~(40nm/5nm) electrode patterns were
first deposited on a thermally
oxidized silicon substrate. It should be noted that 
all the following fabrication processes were
carried out inside a glovebox with an Ar purity of 99.9999\%
since $R$Te$_{3}$ is
extremely sensitive to ambient air~\cite{Kopaczek2023}.
To fabricate thin film devices from the bulk
$R$Te$_{3}$, we used the mechanical exfoliation technique
using scotch tapes. The
exfoliated flakes were transferred from scotch tapes
to transparent polydimethylsiloxane (PDMS) polymers,
and the $R$Te$_{3}$ flakes were released onto the pre-patterned electrodes
aligned under an optical microscope. In addition,
flakes were capped with high quality
hexagonal Boron Nitride (hBN)
to prevent oxidation during taking devices 
from the glovebox and setting them to the measurement system. 
To check the reproducibility, we fabricated at least two different devices for each $R$Te$_{3}$. The main results (obtained with \texttt{\#}1) are shown in the main text, while the additional results (obtained with \texttt{\#}2) are displayed in Supplemental Materials. For simplicity, we omit the device number in the main text.

\begin{figure}%[htbp]
  \begin{center}
    \includegraphics[width = 70mm]{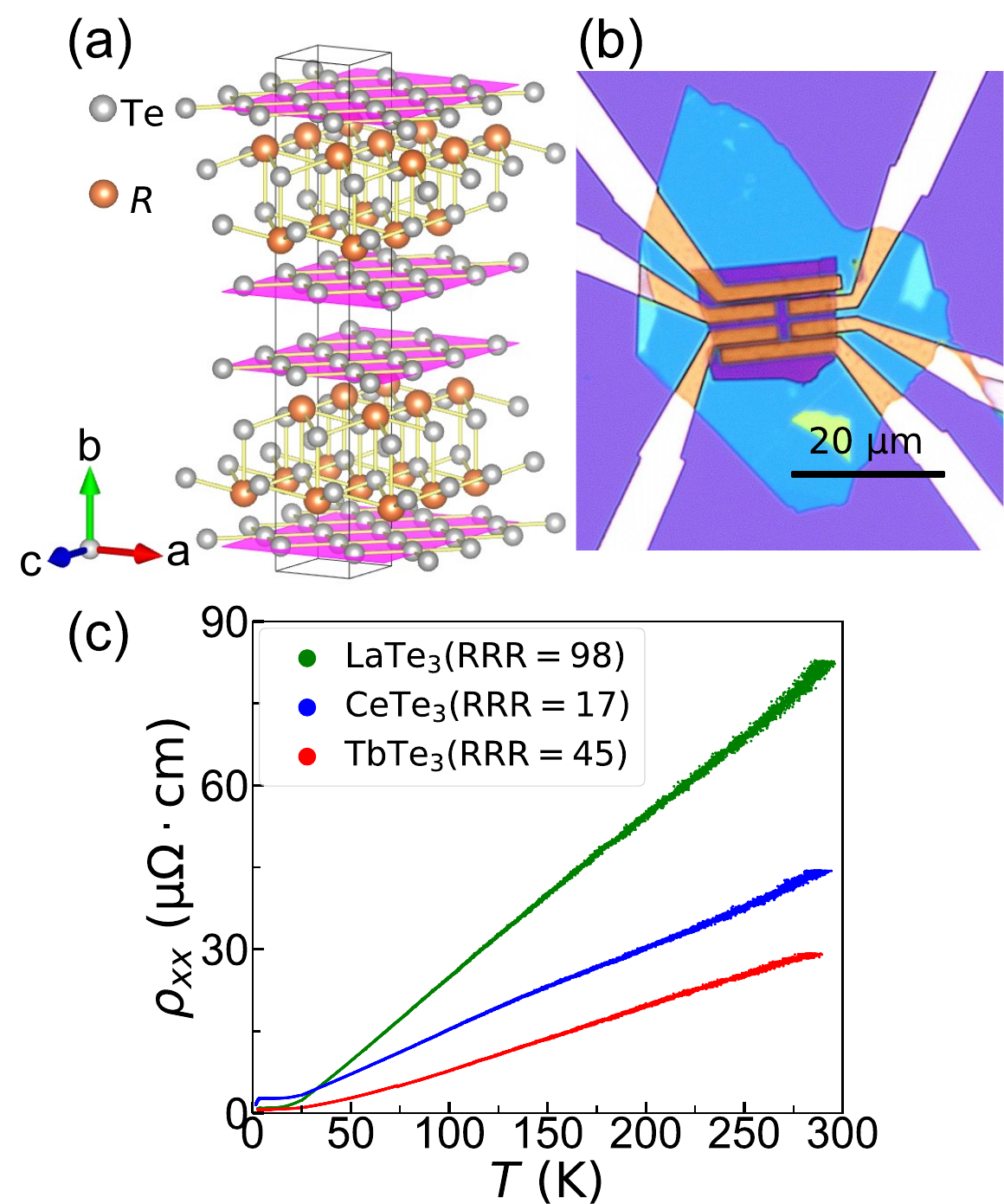}
    \caption{
      (a) Crystal structure of $R$Te$_{3}$. Black line rectangle and purple sheets represent a unit cell and Te square-net sheets, respectively.
      (b) An optical microscope image of a LaTe$_{3}$ thin film device.
      The scale bar corresponds to 20~$\mathrm{\mu m}$. The purple and blue flakes are LaTe$_3$ and hBN, respectively.
      (c) The temperature dependence of the electrical resistivity for LaTe$_{3}$ (green dot), CeTe$_{3}$ (blue dot), and TbTe$_{3}$(red dot) devices.
    }
    \label{fig:1}
  \end{center}
\end{figure}

Figure~\ref{fig:1}(b) shows a typical
device structure taken by an optical
microscope. The purple flake in the center is a 25~$\rm{nm}$ thick
LaTe$_3$ film, while the blue area is the hBN film cap.
Electrical transport measurements
were performed by the conventional four-probe method
with a constant alternating current of 10~$\rm{\mu A}$ or 30~$\rm{\mu A}$ for all devices
using a lock-in amplifier.
The device was cooled with a variable temperature insert
using liquid $^{4}$He down to 1.7~K.
The external magnetic field was
applied using a superconducting magnet up to
8~$\rm{T}$ perpendicular to the $a$-$c$ plane.
The thicknesses of all measured thin films were confirmed by using
an atomic force microscope,
which were 25~nm, 24~nm and 47~nm for LaTe$_{3}$, CeTe$_{3}$,
and TbTe$_{3}$ devices, respectively.
It should be noted that this device fabrication method cannot rule out the possibility 
of a finite strain due to deformation induced in the thin-film devices. 
As detailed in the following section, however, the effect of such a strain would be negligibly small on transport properties
because we did not observe a significant change in the magnetic transition temperature and the temperature dependence of the resistivity, compared to those measured with bulk samples~\cite{Ru2006, Iyeiri2003, Pariari2021, Volkova2022}.

\section{\label{section3}Results and Discussions}
\subsection{\label{section3.1}Longitudinal and Hall resistivities}

The temperature dependence of the longitudinal
resistivity $\rho_{xx}$ for LaTe$_{3}$, CeTe$_{3}$, and TbTe$_{3}$ devices are shown 
in Fig.~\ref{fig:1}(c).
All these show metallic temperature dependencies
even below the CDW transition.
This is due to the partial gap opening,
caused by the imperfect nesting
of the FS~\cite{Brouet2008}.
The residual resistivity ratios (RRRs)
($= \rho_{xx}\left(290\ \rm{K}\right)/\rho_{xx}\left(3.0\ \rm{K}\right)$),
which are indicators of the purity of the
devices, were 98 for LaTe$_{3}$,
17 for CeTe$_{3}$, and 45 for TbTe$_{3}$.
These values are larger than previous studies for TbTe$_{3}$
films~\cite{Xing2020} and comparable to our previous study for
CeTe$_{3}$~\cite{Watanabe2021}.
Although no previous studies have been reported for LaTe$_{3}$
thin films, the RRR values are comparable to the bulk 
counterpart~\cite{Pariari2021}.

\begin{figure*}%[htbp]
  \begin{center}
    \includegraphics[width=\textwidth]{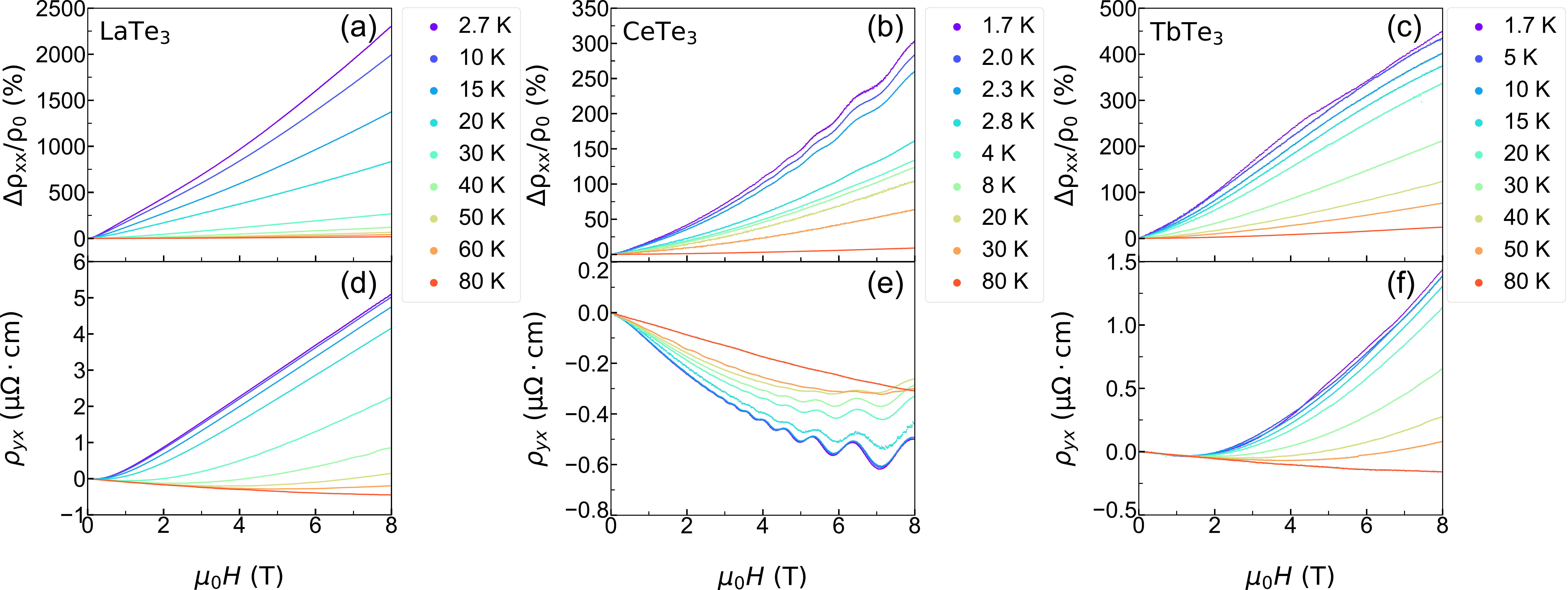}
    \caption{
      (a)-(c) Longitudinal magnetoresistance (MR) ratio $\Delta \rho_{xx}/\rho_{0}$ and (d)-(f) Hall resistivity 
      $\rho_{yx}$ under a magnetic field up to 8~T
      measured at several temperatures for LaTe$_{3}$,
      CeTe$_{3}$, and TbTe$_{3}$ devices, respectively.
    }
    \label{fig:2}
  \end{center}
\end{figure*}

We next performed
longitudinal MR and Hall resistance
measurements at various temperatures from around 100 K down to 1.7 K.
The longitudinal and Hall resistances are symmetrized and antisymmetrized
with respect to the magnetic field
to extract only their respective components.
In Figs.~\ref{fig:2}(a)-\ref{fig:2}(c), we show the magnetic field
dependence of MR ratio defined as 
$\Delta\rho_{xx}/\rho_{0}\equiv \frac{\rho_{xx}(H_z)-\rho_{xx}(H_z=0)}{\rho_{xx}(H_z=0)}$ 
for (a) LaTe$_{3}$, (b) CeTe$_{3}$, and (c) TbTe$_{3}$ devices. 
For all the devices, a non-saturating and large positive MR were observed.
These are characteristics of materials with high carrier 
mobilities $\mu$ ($\mu$ is larger than $\sim 10^3$ cm$^{2}/$V$\cdot$s).
Especially for LaTe$_{3}$,
the MR ratio reached up to 2250\%
at 2.7 K,
which is larger than its bulk counterpart by a rough factor of 
3~\cite{Pariari2021}.
This is likely due to the increase in the crystallinity of the 
thin film device, compared to the bulk counterpart, 
by exfoliation and selection of clean flakes.
In contrast to the parabolic field dependence of MR 
due to the usual Lorentz force contribution in CeTe$_{3}$,
linear MRs were observed
for both LaTe$_{3}$ and TbTe$_{3}$ at low enough temperatures.
These linear MR were also observed
in previous studies of bulk LaTe$_3$ and thin-film
TbTe$_3$ devices~\cite{Pariari2021, Sinchenko2017, Xing2020, Frolov2018}.
Furthermore, in the AFM phase of TbTe$_{3}$ 
below the N\'{e}el temperature ($T_{\rm N1} \approx 6.6$~K), 
the slope of MR is changed at $\mu_{0}H = 3\sim4$~T 
[see $T=1.7$~K in Fig.~\ref{fig:2}(c)]. 
This magnetic field of the change of the MR slope
is close to the phase transition magnetic field reported
in the previous study~\cite{Volkova2022}.
It seems that the AFM
structure has a notable effect 
on the electrical transport properties.

The Hall resistance, which was taken simultaneously 
with the MR, is shown as the Hall resistivity $\rho_{yx}$ 
in Figs.~\ref{fig:2}(d)-\ref{fig:2}(f) 
for (d) LaTe$_{3}$, (e) CeTe$_{3}$, and (f) TbTe$_{3}$ devices.
All the curves show nonlinear external field dependencies.
This behavior was reproducible for all of 
$\rm{LaTe_3}$, $\rm{CeTe_3}$, and $\rm{TbTe_3}$ devices, 
suggesting that these materials possess multiple carriers.
From the Hall curves, we derived the 
carrier mobilities and concentrations, which will be detailed 
in the next subsection.
It should be noted that quantum oscillations were observed 
in the case of CeTe$_3$ both in the longitudinal and 
Hall resistivities. 
The observation of quantum oscillation in the longitudinal MR is 
consistent with our 
previous studies~\cite{Watanabe2021,Watanabe2020}, 
while the emergence of the quantum oscillation in the Hall effect
can be attributed to the multi-carrier effect~\cite{Kikugawa2010}.
The analysis of this SdH signal will be discussed in Sec.~III~C.

\subsection{\label{section3.2}Effect of magnetic order on the electrical transport}
In order to investigate the effect of magnetic order on the electrical transport,
we have analyzed the temperature dependence of the zero field longitudinal 
resistivity, carrier mobility, carrier concentration, 
and MR ratio for each material as shown in Fig.~\ref{fig:3}.
First, we discuss the temperature dependence of the longitudinal resistivity 
in the low temperature region.
In $\rm{CeTe_3}$ and $\rm{TbTe_3}$ which are known to have 
a multi-AFM phase, reductions 
in the zero field resistivity 
were observed below the 
first magnetic transition temperature 
$T_{\rm{N1}}^{\rm{CeTe}_3} \approx 3.0$~K and 
$T_{\rm{N1}}^{\rm{TbTe}_3} \approx 6.6$~K, 
respectively~\cite{Iyeiri2003,Volkova2022}.
Enlarged views of the low-temperature region of
the electrical resistivities of CeTe$_3$ and
TbTe$_3$ with magnetic transition points are shown in
Figs.~\ref{fig:3}(b) and \ref{fig:3}(c), respectively.
The electrical resistivity of nonmagnetic LaTe$_3$ is shown
in Fig.~\ref{fig:3}(a) for comparison, 
where no characteristic reduction was observed.
These reductions in $\rho_{xx}$ for CeTe$_{3}$ and TbTe$_{3}$ 
can be attributed to the suppression of magnetic scattering by magnetic ordering.
It is clear that this effect is especially pronounced in CeTe$_3$ compared to 
TbTe$_3$. This result indicates that 
the Ruderman–Kittel–Kasuya–Yosida (RKKY) interaction plays an important role 
in the magnetism and transport properties in CeTe$_3$.
In fact, the electrical resistivity of CeTe$_3$ also has a small 
bump at around 5 K as shown in Fig.~\ref{fig:3}(b){~\cite{Watanabe2021}. 
This is likely due to
a weak Kondo effect~\cite{Ru2006},
which is further supportive evidence of the AFM coupling 
in the conduction electron of CeTe$_3$.

\begin{figure*}%[htbp]
  \begin{center}
    \includegraphics[width=\textwidth]{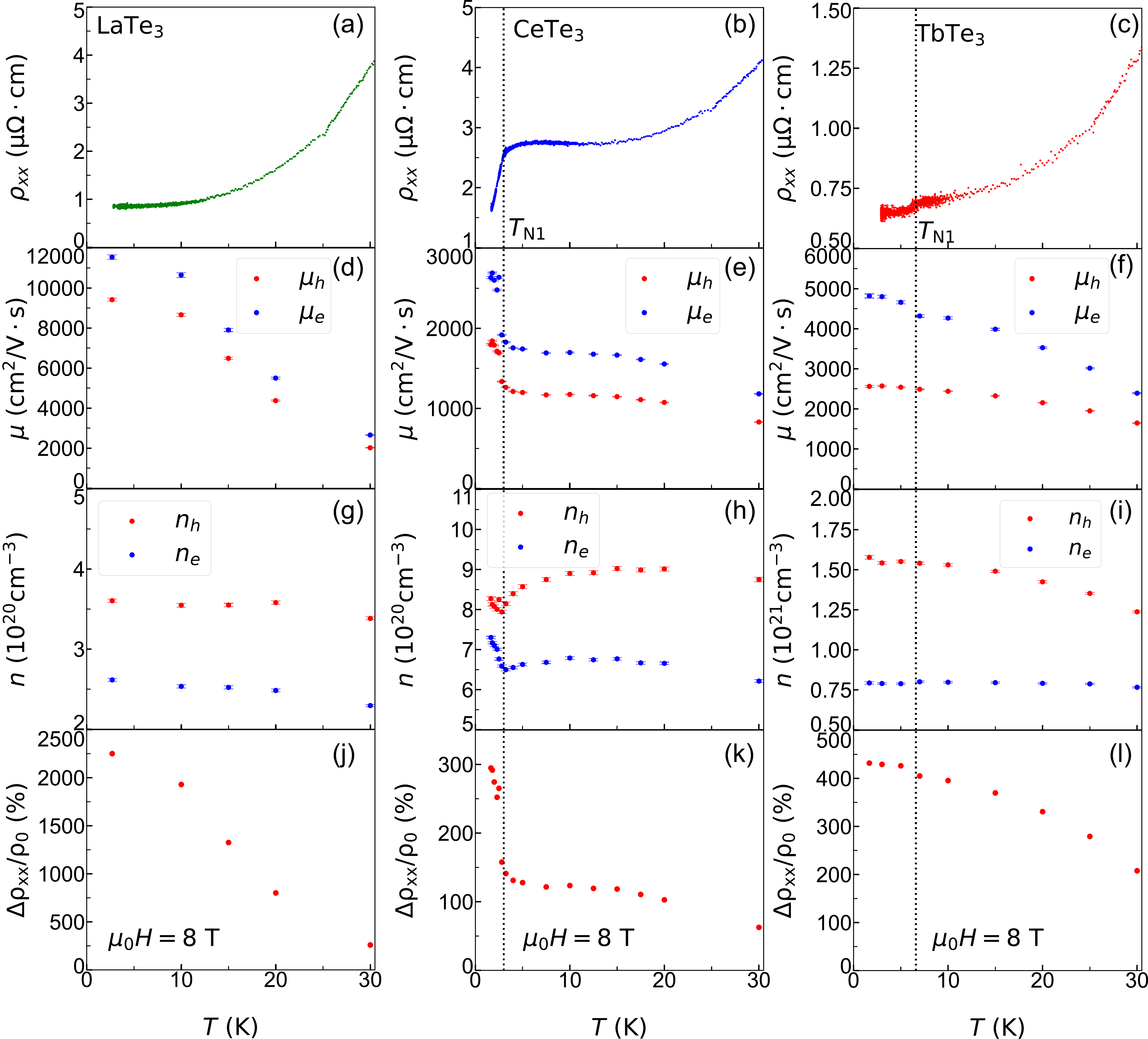}
    \caption{
      (a)-(c) The low-temperature portions of the electrical resistivity of LaTe$_3$, CeTe$_3$, and TbTe$_3$ devices, respectively.
      (d)-(f) The carrier mobilities and (g)-(i) the carrier concentrations of electron and hole obtained
      by the two-band model.
      The error-bar added in the figure indicates the standard deviation errors computed by covariance matrix.
      (j)-(l) The temperature dependence of the MR ratio at 8~T.
    }
    \label{fig:3}
  \end{center}
\end{figure*}

Next we discuss the temperature dependence of the
carrier mobility and concentration calculated from the MR 
and the Hall resistivity.
Assuming the presence of independent electron and hole carriers, 
we have considered the conventional two-band model for analysis and 
used the following equations,
\begin{align}
  \sigma_{xx} & = en_h\mu_h\frac{1}{1+(\mu_hB)^2}+en_e\mu_e\frac{1}{1+(\mu_eB)^2}\label{eq:1}                    \\
  \sigma_{xy} & = \left(n_h\mu_h^2\frac{1}{1+(\mu_hB)^2}-n_e\mu_e^2\frac{1}{1+(\mu_eB)^2}\right)eB\label{eq:2},
\end{align}
where
$\sigma_{xx}=\frac{\rho_{xx}}{\rho_{xx}^2+\rho_{yx}^2}$ and
$\sigma_{xy}=\frac{\rho_{yx}}{\rho_{xx}^2+\rho_{yx}^2}$
are the longitudinal and Hall
conductivities, respectively, $e$ is the elementary charge, 
$B$ is the applied magnetic field, $n_h$ and $n_e$ are 
the carrier concentrations of hole and electron, 
and $\mu_h$ and $\mu_e$ are the carrier mobilities 
of hole and electron, respectively.
Here we have assumed $\rho_{xx}=\rho_{yy}$ and used the Onsager's reciprocal 
relation $\rho_{xy}=-\rho_{yx}$ to derive the electrical conductivity 
from the electrical resistivity.
Carrier mobilities and concentrations 
have been determined by fitting the measured data by the two-band model 
in Eqs.~(\ref{eq:1}) and (\ref{eq:2}) simultaneously.

Although the equations have been well-fitted in most temperatures, 
some discrepancy from the data have been observed at the lower temperatures, 
as shown in Fig.~S1 (in Supplemental Materials).
This could be due to the following three main reasons. 
The first reason is the linear MR; 
while the linear MR was clearly observed, 
the MR described by the usual two-band model is parabolic on magnetic field, 
leading to a difference in fitting. 
The second possibility would be anisotropy in the electrical resistivity;
the assumption $\rho_{xx}=\rho_{yy}$ was used in the conversion from 
electrical resistivity to conductivity, but it is known
that the $R$Te$_3$ system exhibits significant anisotropy 
of electrical resistivity with respect to the crystal axis 
below the CDW transition temperature.
In fact, enhancement of the anisotropy at lower temperatures 
has been reported in some $R$Te$_3$ materials,~\cite{Sinchenko2014,Volkova2022}, 
which could be the cause of errors in the fitting. 
The third possibility would be
multi-carriers; although two-band carriers are
considered here for simplicity, it is likely that more types of carriers 
contribute to transport in reality, and the equation for the two-band model 
may not reflect the actual situation. 
In fact, in the $R$Te$_3$ system, the existence of more than two carriers has been
suggested from measurements of SdH quantum oscillations~\cite{Lei2020,Dalgaard2020}.
In addition, it is possible that the transport
properties unique to the magnetic structure,
such as the anomalous Hall effect, may also
contribute to the discrepancy.

The results of carrier mobilities and concentrations 
for electron and hole obtained by the two-band model 
are shown in Figs.~\ref{fig:3}(d)-\ref{fig:3}(f), 
and Figs.~\ref{fig:3}(g)-\ref{fig:3}(i) 
for LaTe$_3$, CeTe$_3$, and TbTe$_3$ devices, respectively.
The electron mobility of $\rm{LaTe_3}$ is {$\mu_e = 12000$~cm$^2$/V$\cdot$s} at
2.7~K, which is notably high, and the order of magnitude is 
comparable to that of 
GdTe$_3$~\cite{Lei2020},
NdTe$_3$~\cite{Dalgaard2020} and bulk LaTe$_3$~\cite{Pariari2021},
which were reported as high mobility $R$Te$_3$ systems. 
The high mobility common to the $R$Te$_3$ system 
originates from the small pocket created 
by the reconstruction of the FS 
due to the CDW ordering~\cite{Chikina2023}.
In terms of the effect of the magnetic ordering, 
enhancements of carrier mobilities below $T_{\rm{N1}}$
are observed for AFM TbTe$_{3}$ and CeTe$_{3}$. 
As for CeTe$_3$, the carrier mobilities reach up to
a total enhancement of {44\%.}
The only other $R$Te$_3$ material reported to have such a
large mobility enhancement below its magnetic transition temperature
is $\rm{NdTe_3}$~\cite{Dalgaard2020}. 
Similar to the temperature dependence of 
longitudinal resistivity,
the origin of this behavior would be the enhancement of 
electron relaxation time due to the suppression of 
magnetic scattering by magnetic ordering.

The effect of magnetic ordering is also observed 
in the temperature dependence of the MR ratio.
In Figs.~\ref{fig:3}(j)-\ref{fig:3}(l), we have plotted 
the temperature dependence of the MR ratio
at 8~T for (j)~LaTe$_3$, (k)~CeTe$_3$, and (l)~TbTe$_3$ devices.
While a gradual increase in the MR ratio 
with decreasing temperature is observed for the nonmagnetic LaTe$_3$,
discontinuous enhancements of the MR ratio 
below the first magnetic transition temperature
are observed for CeTe$_3$ and TbTe$_3$. In particular, 
a sharp increase of 111\% is observed for CeTe$_3$ below $T_{\rm{N1}}$.
These enhancements of the MR ratio can be explained 
by the increase in the carrier mobility below $T_{\rm{N1}}$. 
On the other hand, we did not observe any negative MR which is often
observed in typical magnetic materials.
This is likely attributed to the fact that 
the positive MR effect 
caused by high carrier mobility overwhelms
the negative MR component.

Let us mention the change in the temperature dependence of carrier concentration below and above $T_{\rm{N1}}$ observed for CeTe$_3$, as shown in Fig.~\ref{fig:3}~(h).
Although the transport phenomena discussed in this section 
can be attributed to the scattering by the fluctuating magnetic moments, 
the change in $n_h$ and $n_e$ cannot be explained
by the same mechanism. Usually, magnetic transition is not accompanied 
by a drastic change in its carrier concentration. 
Rather, we believe that modulation of the CDW,
which is closely related to the band structure of the material, 
could be the direct cause of the carrier concentration temperature dependence. 
Given that this behavior occurs at $T_{\rm{N1}}$,
the data implies that there is some coupling of the CDW and 
magnetic order. 
As far as we know, 
neutron diffraction experiments have been conducted for TbTe$_{3}$~\cite{Pfuner2012, Chillal2020}, where in addition to the commensurate AFM and CDW orders, new magnetic peaks were observed whose propagation vector equals the sum of the AFM and CDW propagation vectors, revealing 
%an entangled relationship. 
a coupling between the orders.
Although there is no evidence for CeTe$_{3}$, the magnetic ordering in CeTe$_3$ may modulate the CDW 
state, resulting in a change of the temperature dependence in $n_h$ and $n_e$.
It is also interesting to note that this effect is pronounced in CeTe$_3$, 
indicating that the RKKY interaction, which is more significant in CeTe$_3$, 
is a possible origin of the coupling between the CDW 
and the magnetic order.

\subsection{\label{section3.3}Shubnikov-de Haas oscillations}

SdH oscillations have 
been reported in several $R$Te$_{3}$ 
materials~\cite{Lei2020,Dalgaard2020,Walmsley2020,Ru2008_2}, 
where the structure of the FS 
plays an important role~\cite{Brouet2008,Chikina2023}.
In this subsection, 
we discuss the effect of magnetic order on the SdH oscillations.
As mentioned in Sec.~III~A.,
in CeTe$_{3}$, SdH oscillations were observed in the MR, i.e., $\rho_{xx}(B)$ [see Fig.~\ref{fig:2}(b)]. In addition, quantum oscillations with the same period as the SdH oscillations were also observed in the Hall resistivity component, i.e., $\rho_{yx}(B)$ [see Fig.~\ref{fig:2}(e)]. In general, SdH oscillations are discussed for $\rho_{xx}(B)$. It is known that the SdH oscillations are also reflected in $\rho_{yx}(B)$ for multi-carrier materials~\cite{Kikugawa2010}. The amplitude of the oscillations in $\rho_{yx}(B)$ becomes larger than that in $\rho_{xx}(B)$.
Thus, we have used the oscillatory component of $\rho_{yx}(B)$ 
for the following analysis of the
quantum oscillations, but as shown in Fig.~S3 (in Supplemental Materials), we have obtained the essentially same result in $\rho_{xx}(B)$.

\begin{figure}%[htbp]
  \begin{center}
    \includegraphics[width=70mm]{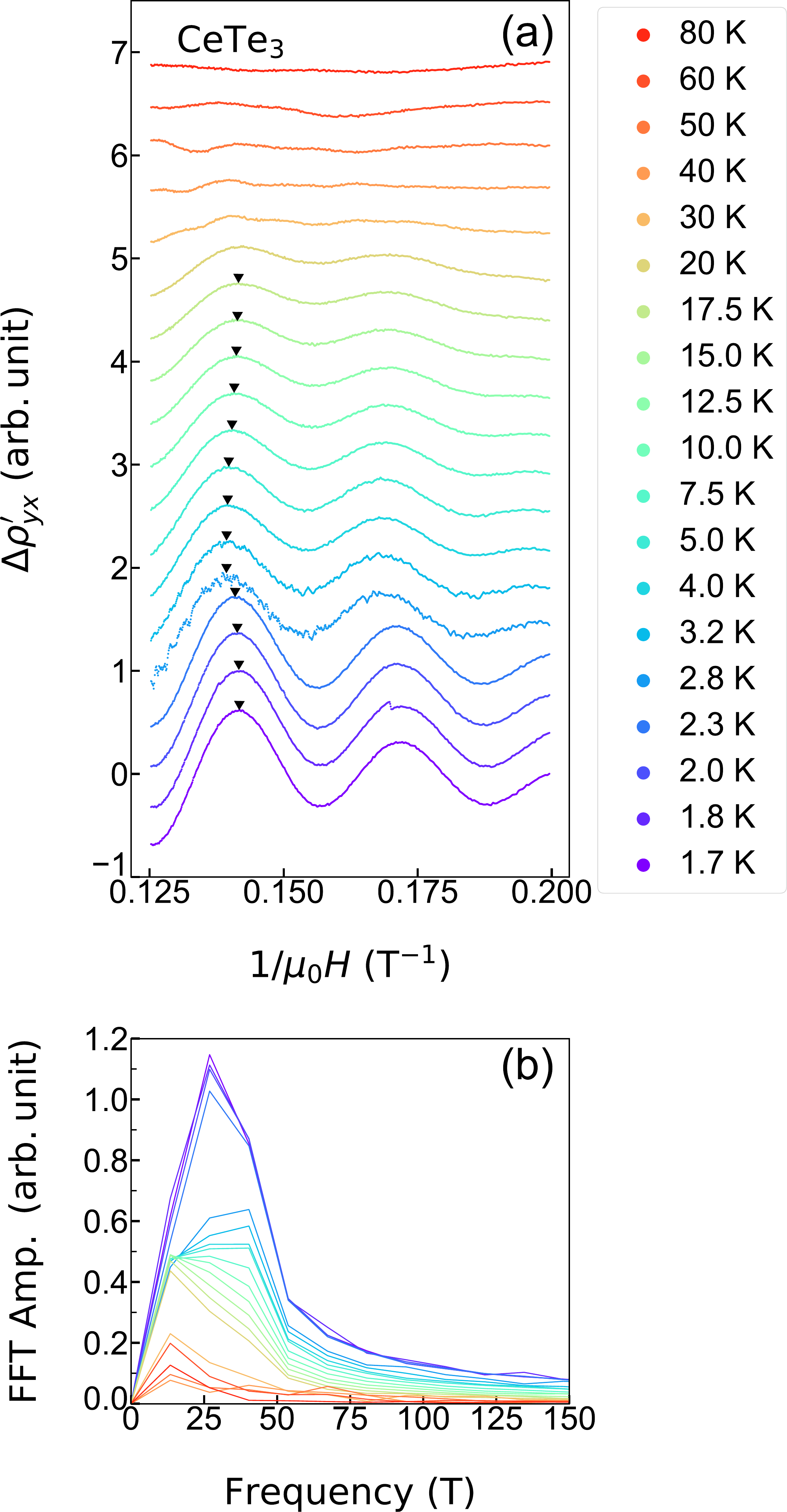}
    \caption{
      (a) SdH oscillation component of CeTe$_{3}$ plotted against the reciprocal of the applied magnetic field for several different temperatures.
      The inverted triangle is an eye guide to show shifts in the positions of the oscillation peaks.
      (b) The fast Fourier transformation (FFT) of the oscillation component for several different temperatures. The correspondence between color and temperature is the same as in (a).
    }
    \label{fig:4}
  \end{center}
\end{figure}

In general, the SdH oscillation is described 
by the following Lifshitz-Kosevich (L-K) formula
with $\Delta\rho'$ as the
oscillating component of resistance, given as
\begin{align}
  \Delta\rho'\propto\frac{\lambda(B)T}{\sinh\left\{\lambda(B)T\right\}}e^{-
  \lambda(B)T_{\rm{D}}}\cos\left\{2\pi\left(\frac{F}{B}-\frac{1}{2}+\beta+\delta\right)
  \right\},
  \label{eq:3}
\end{align}
where $T_{\rm{D}}$ is the Dingle temperature which corresponds to the
blurring of the Landau level, and $F$ is the frequency of the SdH oscillation. 
$2\pi\beta$ is the Berry phase and $\delta$ is the phase shift that 
takes zero in 2D and $\pm\frac{1}{8}$ in 3D systems.
$\lambda(B)$ is defined by $\lambda(B) = 2\pi^2k_{\rm B}m^*/(\hbar eB)$, where $k_{\rm B}$ is 
the Boltzmann constant, $\hbar$ is the reduced 
Planck constant, and $m^*$ is the effective cyclotron mass.

First, we analyze the SdH frequency in order
to determine the FS structure responsible for the
oscillation, using the equation $S = \frac{2\pi eF}{\hbar}$
where $S$ is the extremal surface area of the FS pocket.
For the analysis, the background resistivity 
$\rho_{yx}^{\rm BG}$ was first subtracted from the Hall resistivity 
in the magnetic field range from 5 T to 8 T and 
the the oscillatory component
$\Delta\rho'_{yx} = \rho_{yx} - \rho_{yx}^{\rm BG}$ was obtained
as shown in Fig.~\ref{fig:4}(a). We then performed 
the fast Fourier transformation (FFT) 
to extract the oscillation frequency. 
A single oscillation frequency of $F(\alpha) = 31.8$~T 
was observed as shown in Fig.~\ref{fig:4}(b), which is consistent 
with our previous study~\cite{Watanabe2021}. 
Given that the 3D unit cell of $R$Te$_{3}$ is equal to 
the unit cell of a single Te square-net
rotated by 45 degrees and multiplied by 2~\cite{Brouet2004}, 
the size of the Brillouin zone of CeTe$_{3}$ can be calculated 
as $S_{\rm{BZ}} = \frac{1}{2}\frac{2\pi}{a}\frac{2\pi}{c}$, where
$a = 4.384$~\AA, $c = 4.403$~\AA~\cite{Han2012}.
Therefore, we determine that the measured SdH oscillation originates from a 
FS pocket equivalent to 0.30\% of the entire Brillouin zone, 
which is likely the small FS pocket frequently labeled as 
the $\alpha$ pocket in $R$Te$_{3}$. This pocket arises due to
the reconstruction of the FS under the CDW ordering, 
and was also observed in other $R$Te$_{3}$ 
(0.28\% for LaTe$_3$~\cite{Ru2008_2}, 
0.27\% for GdTe$_3$~\cite{Lei2020}, 
0.2\% for NdTe$_3$~\cite{Dalgaard2020}, and
0.16$\pm$0.1\% for NdTe$_3$ measured by 
angle-resolved photoemission spectroscopy (ARPES)~\cite{Chikina2023}).

\begin{figure}%[htbp]
  \begin{center}
    \includegraphics[width=60mm]{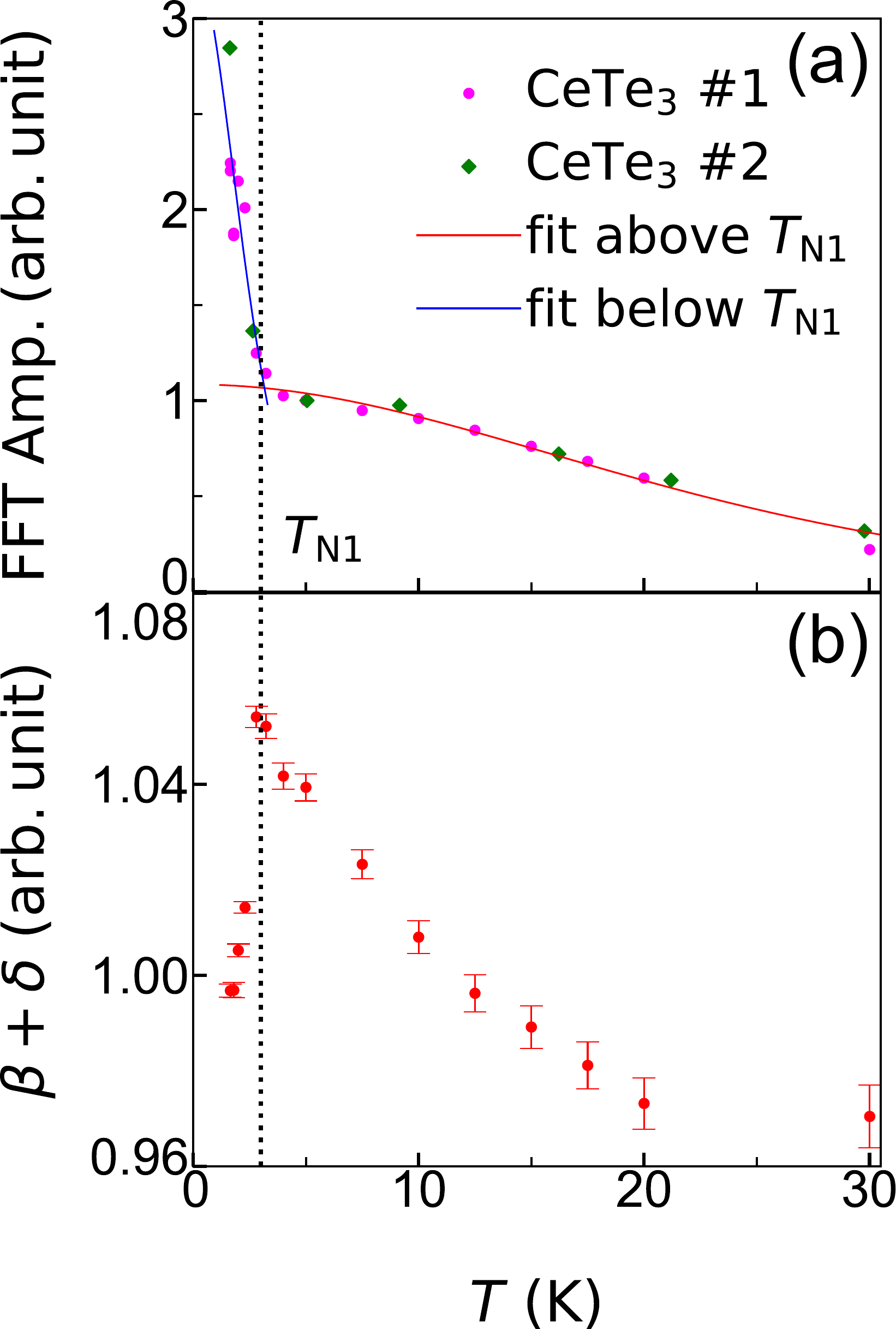}
    \caption{
      (a) The temperature dependence of the FFT amplitudes at 31.8~T with fitting curves of the L-K formula.
      (b) The temperature dependence of $\beta+\delta$.
      The error-bar added in the figure indicates the standard deviation errors computed by covariance matrix.
    }
    \label{fig:5}
  \end{center}
\end{figure}

Next, we discuss the temperature dependence of the SdH oscillation amplitude, 
from which the conduction electron effective mass and mobility can be evaluated.
The SdH oscillation amplitude has a characteristic 
temperature dependence that is proportional to 
$\frac{\lambda(B_0)T}{\sinh\left\{\lambda(B_0)T\right\}}$, 
where $B\rm{_0}$ is the mean value of the analyzed magnetic field 
range. By fitting the FFT amplitude 
with the above function as shown in Fig.~\ref{fig:5}(a),
the effective mass $m^*$ can be obtained.
We note that data from multiple samples are plotted in Fig.~\ref{fig:5}(a),
which are normalized by the FFT amplitude at 5~K.
The results are listed in Table~\ref{table:1}. 
Above $T_{\rm N1}$, the effective mass is 0.042$m_{e}$, which is 
comparable to our previous work~\cite{Watanabe2021}. 
A large deviation from the L-K formula 
has been observed below $T_{\rm{N1}}$. 
Thus, we performed separate fits for the effective mass 
above and below $T_{\rm{N1}}$, resulting in a large 
enhancement of effective mass (0.37$m_{e}$) below $T_{\rm{N1}}$.
The quantum life time $\tau_q = \hbar/(2\pi k_{\rm B} T_{\rm{D}})$ 
and the carrier mobility $\mu_q = e\hbar/(2\pi k_{\rm B}m^*T_{\rm{D}})$ 
can also be evaluated as listed in Table~\ref{table:1}.
We have observed an enhancement 
in all $m*$, $\tau_q$ and $\mu_q$ below $T_{\rm{N1}}$.
It is important to note that these enhancements provide further supportive 
evidence of our discussion based on the MR and Hall resistivity 
in Sec.~III~B.
The enhancement of $m^*$ below $T_{\rm{N1}}$ is most likely 
attributed to some modulation of the band structure near the FS 
from the magnetic order, which is consistent with 
our analysis of the temperature dependence of 
the carrier concentrations.
The enhancement of carrier mobility is 
also consistent with the temperature dependence of 
longitudinal and Hall resistivities. 
It should originate from the increase in relaxation time 
of the carriers 
as the fluctuations of magnetic moments are suppressed below $T_{\rm{N1}}$.

Finally, we point out the $\beta + \delta$ term in Eq.~(\ref{eq:3}). 
As mentioned, the $\beta$ and $\delta$ terms 
correspond to the Berry curvature
and the dimension of the FS pocket, respectively. 
The temperature dependence of $\beta + \delta$ is shown 
in Fig.~\ref{fig:5}(b), where a peak structure
is observed at $T_{\rm{N1}}$.
This phase shift can also be seen in the shift of the position of the oscillation peaks shown in Fig.~\ref{fig:4}(a).
Although further study is required to elucidate the
origin of this behavior, it is another supportive data 
which indicates the modulation of the band structure 
near the FS around $T_{\rm{N1}}$, since this quantity should be sensitive to 
the FS topology.

\begin{table}%[htbp]
  \centering
  \footnotesize
    \caption{Comparison of effective cyclotron mass, Dingle temperature, scattering relaxation time, and carrier mobility for PM and AFM regions of CeTe$_{3}$.}\label{table:1}
  \begin{tabular}{ccccc}
    \hline \hline
    region & $m^*/m_e$ & $T_{\rm{D}}{\rm(K)}$  & $\tau_q\rm{(s)}$  & $\mu_q\rm{(cm^2/Vs)}$ \\
    \hline
    PM (10~K)                            & 0.042             & 48.3                  & $2.51\times10^{-14}$ & 1056                  \\
    AFM (1.7~K)                           & 0.37              & 2.53                  & $4.80\times10^{-13}$ & 2266                  \\
    \hline \hline
  \end{tabular}
\end{table}

\section{\label{section4}Conclusion}
In this work, we have focused on $R$Te$_{3}$, 
a van-der-Waals material with high mobility. 
We performed systematic longitudinal resistivity and Hall resistivity measurements 
on single-crystal thin films of nonmagnetic LaTe$_{3}$ and
antiferromagnetic CeTe$_{3}$ and TbTe$_{3}$.
We have observed a peculiar linear magnetoresistance 
in LaTe$_3$ and TbTe$_3$, the magnetic field dependence of 
nonlinear Hall resistance in all samples, and 
a SdH oscillation in CeTe$_{3}$. 
Carrier concentration and mobility are derived using a two band model. 
Both CeTe$_3$ and TbTe$_3$ show decreases in longitudinal resistivity 
and increases in carrier mobility below $T_{\rm{N1}}$. 
Combined with the quantum mobility obtained from the SdH oscillation in CeTe$_{3}$, 
we conclude that the behavior is caused by the suppression of 
magnetic scattering due to the magnetic ordering. 
Furthermore, a change in the temperature dependence of the carrier
concentration is also observed in CeTe$_3$ below and above its magnetic transition temperature. Combined with the results
of L-K formula fitting, we have found that the band structure near the FS 
may be modulated by the magnetic ordering. 
This indicates that the magnetic ordering and CDW ordering 
are coupled via the RKKY interaction.
It is expected that further ARPES experiments before and after the magnetic 
transition will be performed to search for interesting properties 
in the system where CDW and magnetic order are strongly coupled.
Our results demonstrates not only possibilities of two-dimensional devices 
with high mobility and magnetic order, but also a test-bed towards 
the understanding of new phenomena due to the coexistence of magnetic order 
and CDW order in low-dimensional materials.

\section{Data Availability}
The data that support the findings of this study are available
from the corresponding author upon reasonable request.

\section{Acknowledgements}
We thank S. Nakaharai for his technical supports and fruitful discussions.
This work was supported by JSPS KAKENHI (Grant 
Nos. JP20H02557, JP21J20477, JP22H04481, JP23H00257, and JP23K13062) and 
JST FOREST (Grant No. JPMJFR2134).

\end{document}